\documentclass[aps,preprint]{revtex4}

\usepackage{graphicx}

\newcommand{\be}{\begin{equation}}
\newcommand{\ee}{\end{equation}}

\begin{document}

\title{\bf Spin polarized liquid $^3{\rm He}$
}
\author{G.H. Bordbar\footnote{Corresponding author}
\footnote{E-Mail: bordbar@physics.susc.ac.ir}, S.M. Zebarjad, M.R.
Vahdani and M. Bigdeli}
 \affiliation{Department of Physics, Shiraz
University, Shiraz 71454, Iran}
\begin{abstract}
We have employed the constrained variational method to study the
influence of spin polarization on the ground state properties of
liquid $^3{\rm He}$. The spin polarized phase, we have found, has
stronger correlation with respect to the unpolarized phase. It is
shown that the internal energy of liquid $^3{\rm He}$ increases by
increasing polarization with no crossing point between polarized
and unpolarized energy curves over the liquid density range. The
obtained internal energy curves show a bound state, even in the
case of fully spin polarized matter. We have also investigated the
validity of using a parabolic formula for calculating the energy
of spin polarized liquid $^3{\rm He}$. Finally, we have  compared
our results with other calculations.
\end{abstract}
\maketitle
\newpage
\section{Introduction}
\label{intro} The spin polarized liquid $^3{\rm He}$ is an
interesting quantum many-body system which can be experimentally
examined. In fact, it is expected that this phase of liquid
$^3{\rm He}$ has a large life time to be observed in a
quasi-thermodynamic equilibrium \cite{1,1p,1p1,1p2,2,3}.

Some theoretical investigations have been done for spin polarized
liquid $^3{\rm He}$ using different approaches such as Green's
function Monte Carlo (GFMC), Fermi hyper-netted  chain (FHNC),
correlated basis functions (CBF) and transport theory
\cite{4,4p,5,6,6p,7,8}. Recently, we have studied the unpolarized
liquid $^3{\rm He}$ and calculated some of its thermodynamic
properties at finite temperature. A good agreement between our
results and corresponding empirical values has been shown
\cite{9,10}. In these calculations, constrained variational method
based on the cluster expansion of the energy functional has been
used. This method is a powerful microscopic technique used in
many-body calculations of dense matter
\cite{11,12,13,14,15,16,17}. In this article, we employ this
method to investigate the ground state of spin polarized liquid
$^3{\rm He}$.
\section{Constrained Variational Calculation of Spin
Polarized Liquid $^3{\rm He}$} To calculate the ground state
energy of spin polarized liquid $^3{\rm He}$, consisting of $N$
interacting atoms with $N^{(+)}$ spin up and $N^{(-)}$ spin down
atoms, $N=N^{(+)}+N^{(-)}$, we use the variational calculation
based on the cluster expansion of the energy  functional
\cite{18}. We consider  up to the two-body energy term in the
cluster expansion,
\begin{equation}\label{eq1}
  E=E_1+ E_2.
\end{equation}
The one-body energy per particle for the spin polarized matter is
given by
\begin{eqnarray}\label{eq2}
  E_1&=&E_1^{(+)}+E_1^{(-)}\nonumber\\
  &=&\frac{3}{10}\bigg(\frac{\hbar^2}{2m}\bigg)(3\pi^2)^{2/3}
 \bigg[ (1+\xi)^{5/3}+(1-\xi)^{5/3}\bigg]
  \rho^{2/3},
\end{eqnarray}
where  $\rho$ is the total number density,
\begin{equation}\label{eq4}
\rho=\rho^{(+)}+\rho^{(-)},
\end{equation}
and the  spin asymmetry parameter, $\xi$, is defined to be
\begin{equation}\label{eq3}
\xi=\frac{N^{(+)}-N^{(-)}}{N}.
\end{equation}
$\xi$ shows the spin ordering of the matter which can get
a value in the range of $\xi=0.0$
(unpolarized matter) to $\xi=1.0$ (fully polarized matter).

To obtain the two-body energy term, $E_2$, we can start from the
known equation
\begin{equation}\label{eq5}
 E_2=\frac{1}{2}\sum_{ij}<ij|W(12)|ij>_a,
\end{equation}
where
\begin{equation}\label{eq6}
W(12)= \frac{\hbar^2}{m}\bigg(\nabla f(12)\bigg)^2+f^2(12)V(12),
\end{equation}
and $V(12)$ is the two-body potential between the helium atoms (we
use the Lennard-Jones potential with $\epsilon =10.22~K$ and
$\sigma =2.556~A$) and $f(12)$ is two-body correlation function.
By considering the $|i>$ as a plane wave and performing some
algebra, we have derived the following relation for the spin
polarized matter,
\begin{equation}\label{eq7}
E_2=\frac{1}{2}\rho^2\int dr_1\int dr_2\bigg\{
1-\frac{1}{4}\bigg[(1+\xi)^2 \ell^2(k_F^{(+)} r_{12})+(1-\xi)^2
\ell^2(k_F^{(-)} r_{12})\bigg]\bigg\}W(r_{12}),
\end{equation}
where
\begin{equation}\label{eq8}
\ell(x^{(i)})=\frac{3}{(x^{(i)})^3}\bigg[ \sin(x^{(i)})-x^{(i)}
\cos(x^{(i)})\bigg],
\end{equation}
and $x^{(i)}$ is $k_F^{(+)} r_{12}$ or $k_F^{(-)} r_{12}$.
$k_F^{(+)}=(6\pi^2\rho^{(+)})^{1/3} $ and $k_F^{(-)}
=(6\pi^2\rho^{(-)})^{1/3}$ are the Fermi momentum of spin up and
spin down states, respectively. By minimizing the Eq.~(\ref{eq7})
with respect to the $f(12)$, the following differential equation
is obtained:
\begin{equation}\label{eq9}
\frac{d}{dr}\bigg[L(r)f^\prime(r)\bigg]
-\frac{m}{\hbar^2}\bigg[V(r) +\lambda\bigg]
L(r)f(r) =0,
\end{equation}
where
\begin{equation}\label{eq10}
L(r)= 1-\frac{1}{4}\bigg[(1+\xi)^2 \ell^2(k_F^{(+)} r_{12})+(1-\xi)^2
\ell^2(k_F^{(-)} r_{12})\bigg],
\end{equation}
and $\lambda$ is the lagrange multiplier which imposes the
constraint of the two-body wave function normalization.
Eq.~(\ref{eq9}) can  be solved numerically and the correlation
function $f(12)$ and finally the internal energy of  system are obtained.
\section{Results}
\begin{figure}
\includegraphics[height=15cm]{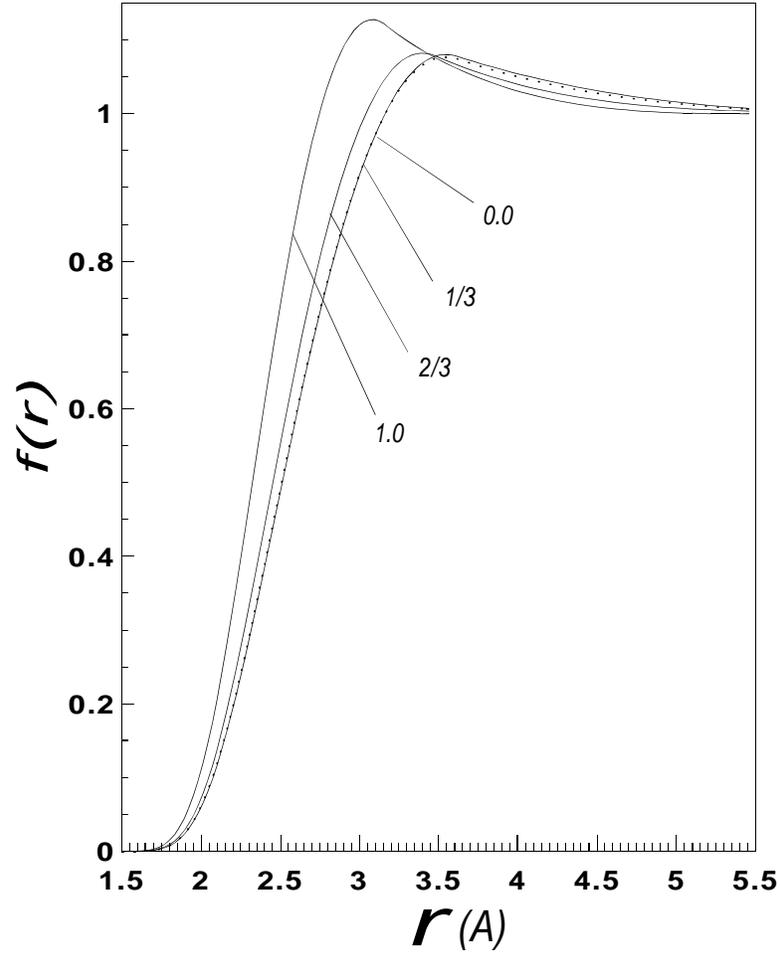}
\caption{ Two-body  correlation function of liquid ${\rm ^3He}$,
$f(r)$, as a function of interatomic distance ($r$) at different
values of spin asymmetry parameter  $\xi =$ 0.0, 1/3, 2/3 and 1.0
for $\rho=0.01~ \rm{A^{-3}}$.} \label{fig1}
\end{figure}

The numerical results for the two-body correlation function of
liquid ${\rm ^3He}$ are given in Fig.~\ref{fig1}.
This shows that  for the higher value of spin
asymmetry parameter, the correlation function increases more
rapidly and reaches to the limiting value ($f(r)=1$) at smaller value
of $r$. From Fig.~\ref{fig1}, we can see that
in the case of higher polarization, the  helium
atoms have a stronger correlation at short relative distance.

\begin{figure}
\includegraphics[height=15cm]{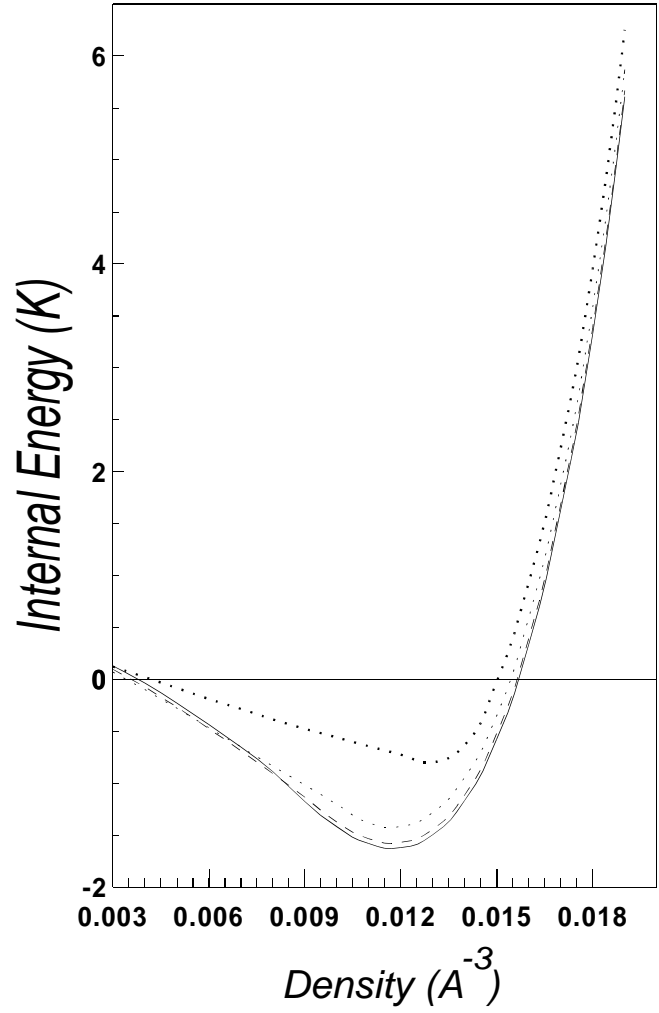}
\caption{Internal energy of liquid ${\rm ^3He}$ versus total
number density at $\xi =$ 0.0 (full curve), 1/3 (dashed curve),
2/3 (dotted curve) and 1.0 (heavy dotted curve).} \label{fig2}
\end{figure}

The numerical values of kinetic, potential and internal energy of
liquid ${\rm ^3He}$ for the different polarizations versus total
number density are presented in Tables~\ref{tab1}-\ref{tab4}. Our
results of internal energy for different spin asymmetry parameters
are compared in Fig.~\ref{fig2}. This figure indicates that as
polarization increases, the internal energy gets the higher values
over the liquid density range and there is no crossing point
between the internal energy curves of polarized and unpolarized
cases in this region. This behavior is in agreement with
experiment \cite{6,7,8}. A comparison between Tables~\ref{tab1},
\ref{tab2}, \ref{tab3} and \ref{tab4} shows that for each density,
the kinetic energy (potential energy) increases (decreases) by
increasing spin asymmetry parameter. Over the liquid density
range, the increasing of kinetic energy dominates which leads to
the higher internal energies. From Fig.~2, it can be also seen
that for all values of $\xi$, the energy curve has a minimum which
shows the existence of a bound state for this system, even for the
fully polarized matter ($\xi=1.0$). We see that by increasing the
spin asymmetry parameter,  this minimum point of energy curve
shifts to the higher densities.

\begin{figure}
\includegraphics[height=15cm]{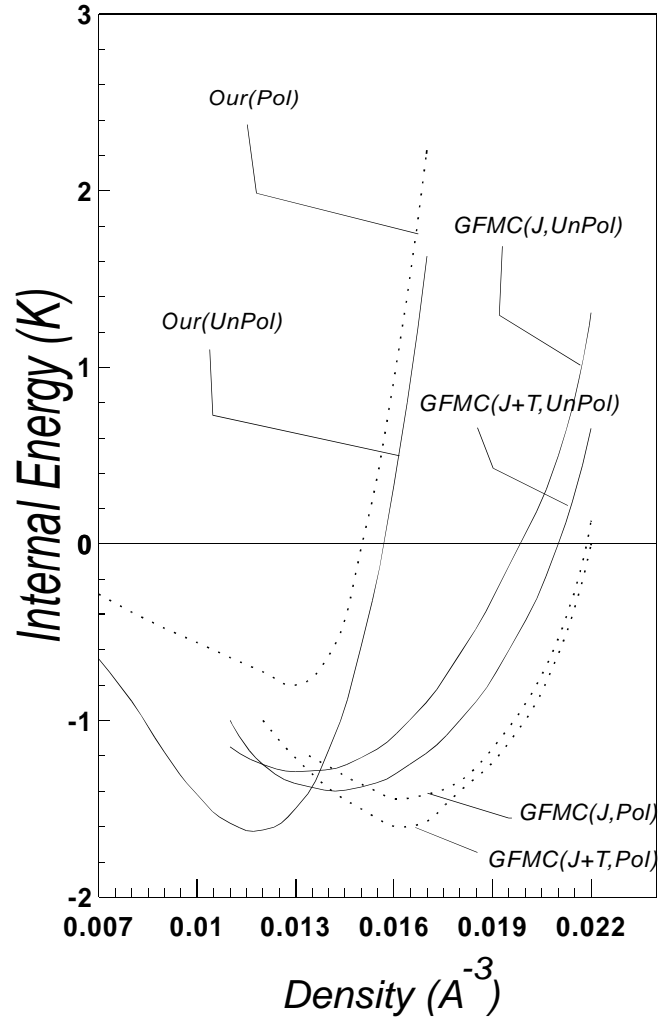}
\caption{Comparison of our results for the unpolarized (UnPol) and
fully polarized (Pol) cases with the results of GFMC for different
choices of the wave function. J refers to Jastrow, and J+T refers
to Jastrow plus three-body wave functions~\cite{7}.} \label{fig3}
\end{figure}

\begin{figure}
\includegraphics[height=15cm]{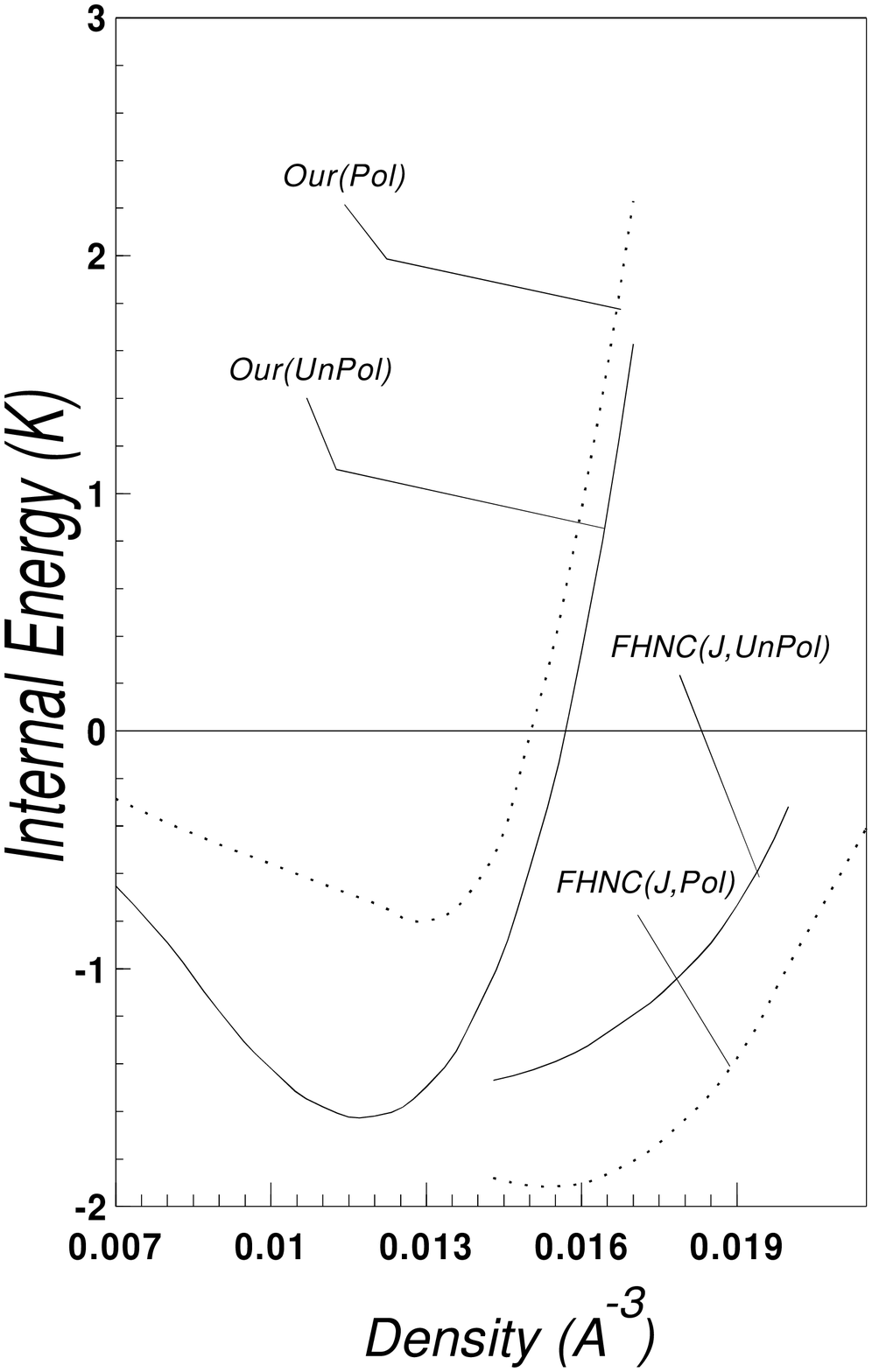}
\caption{Comparison of our results for the unpolarized (UnPol) and
fully polarized (Pol) cases with the results of FHNC~\cite{6}. J
refers to Jastrow wave function. } \label{fig4}
\end{figure}

\begin{figure}
\includegraphics[height=15cm]{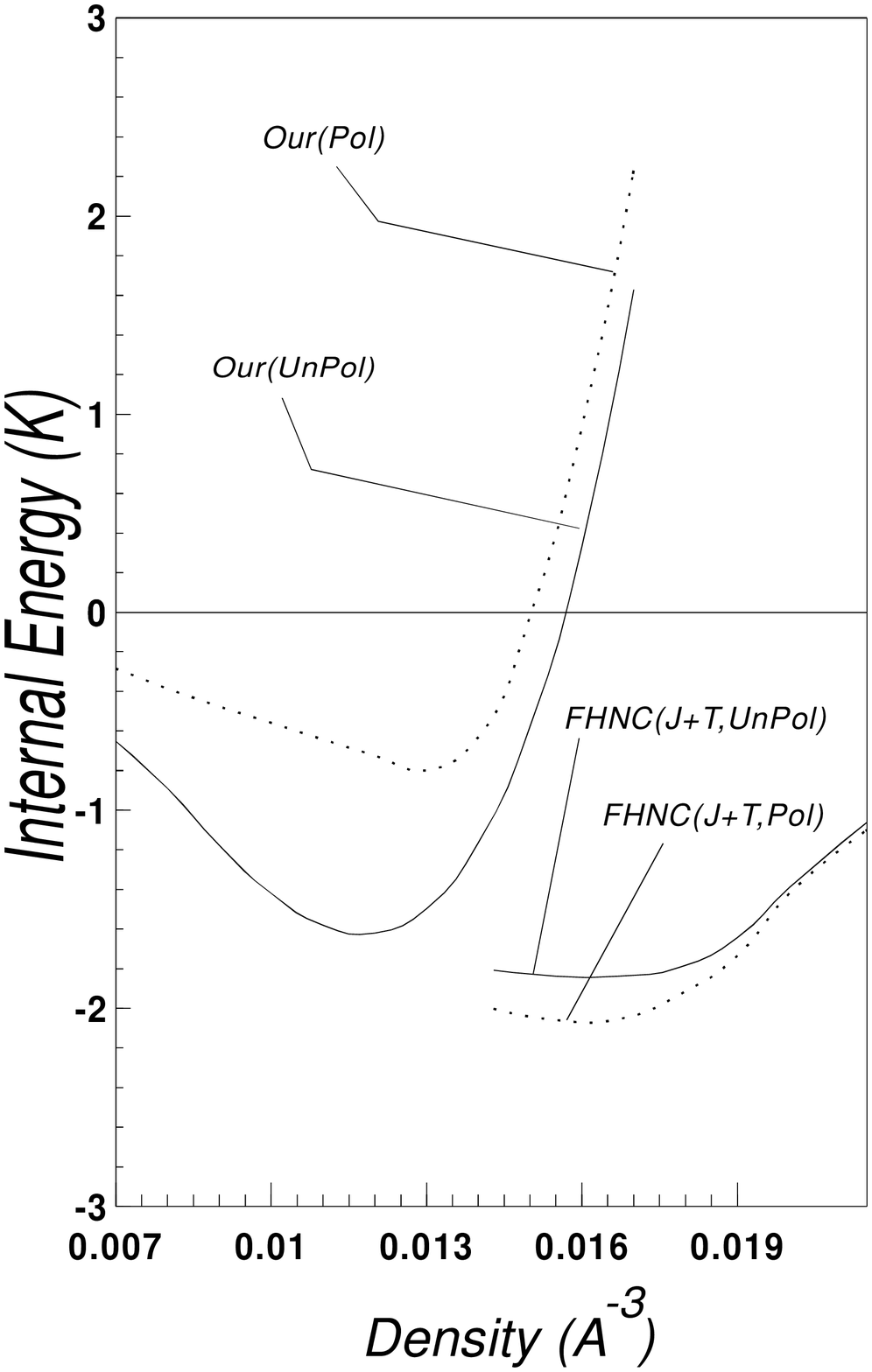}
\caption{ Comparison of our results for the unpolarized (UnPol)
and fully polarized (Pol) cases with the results of FHNC~\cite{6}.
J+T refers to Jastrow  plus three-body wave function.}
\label{fig5}
\end{figure}

\begin{figure}
\includegraphics[height=15cm]{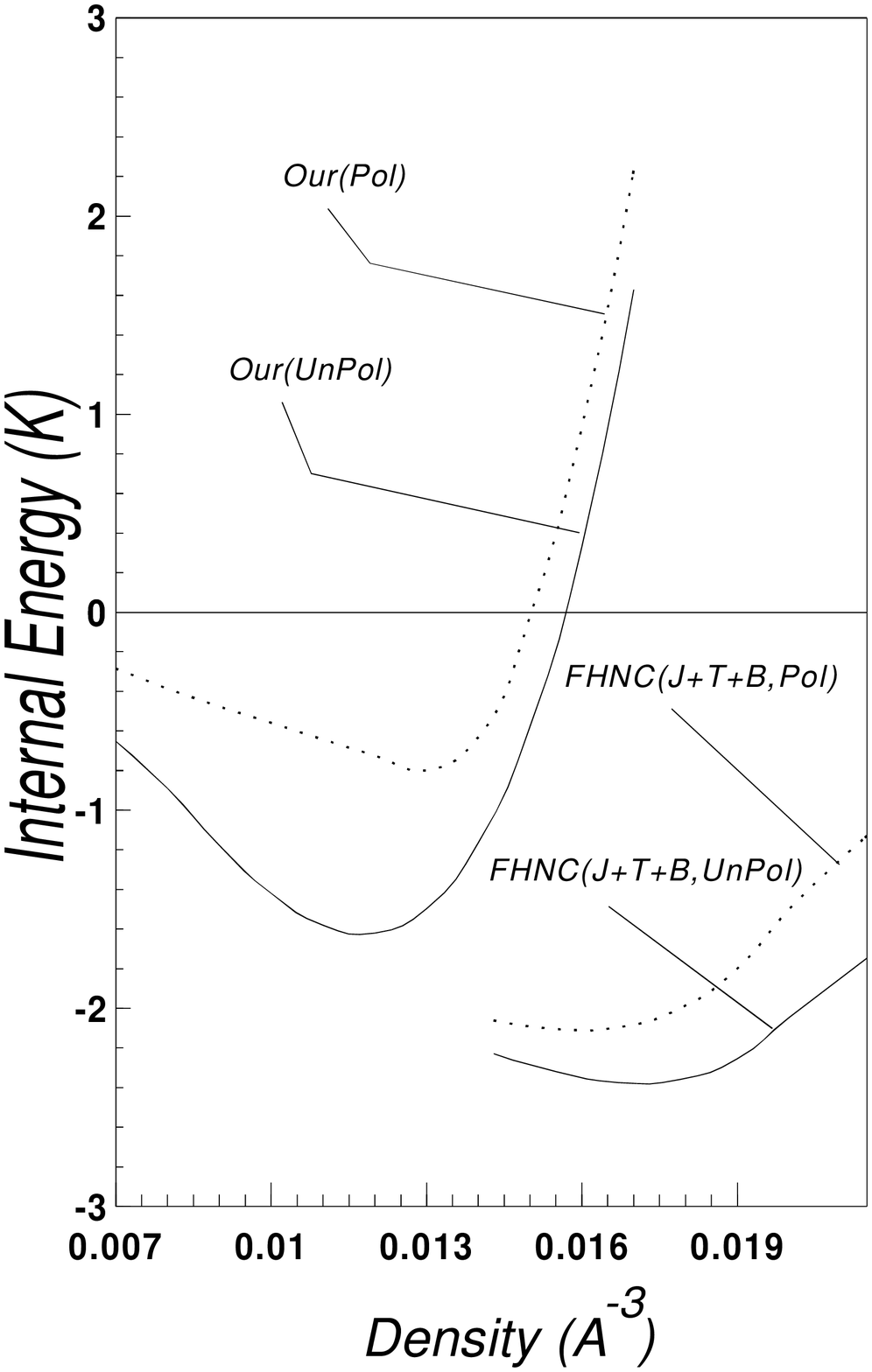}
\caption{ Comparison of our results for the unpolarized (UnPol)
and fully polarized (Pol) cases with the results of FHNC~\cite{6}.
J+T+B refers to Jastrow  plus three-body plus backflow wave
function.} \label{fig6}
\end{figure}

To compare our method with the well-known many-body  techniques,
we present the results of Green's function Monte Carlo (GFMC) and
Fermi hyper-netted chain (FHNC) calculations in
Figs.~\ref{fig3}-\ref{fig6}. In Fig.~\ref{fig3}, the internal
energy of fully polarized and unpolarized liquid ${\rm ^3He}$
calculated
  with GFMC method \cite{7,8} have been compared with our results.
There are two points that we can mention from this figure. First,
the crossing point problem exist and secondly, for all densities
approximately  grater than $0.014 A^{-3}$, the energy of polarized
case is lower than unpolarized case which are not acceptable from
the experiment. However, these problems do not exist in our
method, although we should include the three-body correlation
effect to obtain a better result. In Figs.~\ref{fig4}-\ref{fig6},
the results of FHNC method for different choices of wave
function~\cite{6} are presented. We can see from these figures
that before considering the backflow effect (momentum dependent
two-body correlation), the polarized curve is always lower than
the unpolarized curve. However after performing extra
calculations, the appropriate results have been obtained in
Fig.~\ref{fig6}. This shows that our constrained variational
method can do the job much simpler, although we still need to add
the  three-body cluster energy to obtain a better
result~\cite{17,19}.

 There is a similarity between spin asymmetry parameter in our
calculations and isospin asymmetry parameter in nuclear matter
calculations in which the energy of asymmetrical nuclear matter
can be calculated using the parabolic approximation. In this
approximation, one considers only the quadratic term in asymmetry
parameter as well as the energy of symmetric matter \cite{16}. In
a similar way, we can define the following relation for the
internal energy of spin polarized liquid $\rm{^3He}$:
\begin{equation}\label{eq11}
 E(\rho,~\xi)=  E(\rho,~\xi=0.0)+a_{asym.}(\rho)~\xi^2,
\end{equation}
where the above equation gives the definition of spin asymmetry
energy $a_{asym.}$ as:
\begin{equation}\label{eq12}
a_{asym.}(\rho)= E(\rho,~\xi=1.0)-  E(\rho,~\xi=0.0).
\end{equation}
 $ E(\rho,~\xi=0.0)$ is the internal
 energy of unpolarized matter which
has a symmetric configuration in spin state and $E(\rho,~\xi=1.0)$
is the internal energy of fully polarized matter.
It is now interesting to
see if this  approximation, Eq.~(\ref{eq11}),   agrees with our
microscopic calculations. For this purpose, we have compared the
internal energy of spin polarized liquid $^3{\rm He}$ at different
$\xi$ for  both microscopic calculation and using parabolic
approximation in Tables~\ref{tab5} and \ref{tab6}. These tables
indicate an agreement between these two approaches, specially at
high densities.

\section{Summary and Conclusion}
We have considered a system consisting of $N$ Helium atoms
($^3{\rm He}$) with an asymmetrical spin configuration and derived
the two-body term in the the cluster expansion of the energy
functional. Then, we have minimized  the two-body energy term
under the normalization constraint and obtained the differential
equation. The numerical results of internal energy of this system
have been presented for different values of  spin asymmetry
parameter and density. It is found that as the polarization of
liquid $^3{\rm He}$ increases, the two-body correlation becomes
stronger. Over the liquid density range, our results show that the
internal energy of liquid $^3{\rm He}$ increases by increasing
spin asymmetry parameter with no crossing point between polarized
and unpolarized energy curves. This shows an agreement with
experimental results. It is also seen that there is a bound state
for all values of polarization. The validity of using parabolic
approximation in calculating the energy of spin polarized matter
is shown. Therefore, by using this approximation one can preform
calculations for the spin polarized matter much simpler. we have
also compared our results with other calculations to show that why
of our  constrained variational method is a powerful technique.

\acknowledgements {
 Financial support from Shiraz University
research council is gratefully acknowledged.}

\newpage
\begin{table}
\begin{center}
\caption{Kinetic energy (KE), potential energy (PE) and internal
energy of unpolarized liquid $^3{\rm He}$ ($\xi = 0.0$) versus
total number density.} \label{tab1}
\begin{tabular}{c|c|c|c}
\hline
Density (${\rm A^{-3}}$) & KE (K) & PE (K) & Internal Energy (K)\\
\hline
  0.003&      0.960  &   -0.832&      0.128\\
  0.005&      1.349  &   -1.578&     -0.228\\
  0.007&      1.689  &   -2.341&     -0.651\\
  0.009&      1.997  &   -3.176&     -1.178\\
  0.011&      2.283  &   -3.864&     -1.581\\
  0.013&      2.552  &   -4.049&     -1.496\\
  0.015&      2.807  &   -3.386&     -0.578\\
  0.017&      3.052  &   -1.422&      1.629\\
  0.019&      3.287  &    2.321&      5.608\\
\hline
\end{tabular}
\end{center}
\end{table}
\newpage
\begin{table}
\begin{center}
\caption{As Table~\ref{tab1}, but for spin polarized liquid
$^3{\rm He}$ at $\xi=1/3$.} \label{tab2}
\begin{tabular}{c|c|c|c}
\hline
Density (${\rm A^{-3}}$) & KE (K) & PE (K) &Internal energy (K)\\
\hline
  0.003&      1.019&     -0.922&      0.096\\
  0.005&      1.433&     -1.701&     -0.268\\
  0.007&      1.794&     -2.482&     -0.688\\
  0.009&      2.121&     -3.256&     -1.135\\
  0.011&      2.424&     -3.957&     -1.532\\
  0.013&      2.710&     -4.154&     -1.444\\
  0.015&      2.981&     -3.505&     -0.523\\
  0.017&      3.241&     -1.555&      1.685\\
  0.019&      3.491&      2.176&      5.667\\
\hline
\end{tabular}
\end{center}
\end{table}
\newpage
\begin{table}
\begin{center}
\caption{As Table~\ref{tab1}, but for spin polarized liquid
$^3{\rm He}$ at $\xi=2/3$.} \label{tab3}
\begin{tabular}{c|c|c|c}
\hline
Density (${\rm A^{-3}}$) & KE (K) & PE (K) &Internal energy (K)\\
\hline
  0.003&      1.202&     -1.133&      0.069\\
  0.005&      1.689&     -1.968&     -0.279\\
  0.007&      2.114&     -2.766&     -0.652\\
  0.009&      2.499&     -3.518&     -1.018\\
  0.011&      2.857&     -4.240&     -1.382\\
  0.013&      3.194&     -4.477&     -1.283\\
  0.015&      3.514&     -3.863&     -0.349\\
  0.017&      3.819&     -1.945&      1.874\\
  0.019&      4.114&      1.754&      5.868\\
\hline
\end{tabular}
\end{center}
\end{table}
\newpage
\begin{table}
\begin{center}
\caption{As Table~\ref{tab1}, but for fully polarized liquid
$^3{\rm He}$ ($\xi=1.0$).} \label{tab4}
\begin{tabular}{c|c|c|c}
\hline
Density (${\rm A^{-3}}$) & KE (K) & PE (K) & Internal energy (K)\\
\hline
  0.003&      1.524&     -1.403&      0.120\\
  0.005&      2.142&     -2.227&     -0.085\\
  0.007&      2.681&     -2.967&     -0.286\\
  0.009&      3.170&     -3.648&     -0.478\\
  0.011&      3.624&     -4.268&     -0.644\\
  0.013&      4.051&     -4.850&     -0.799\\
  0.015&      4.457&     -4.478&     -0.021\\
  0.017&      4.845&     -2.615&      2.229\\
  0.019&      5.218&      1.033&      6.251\\
\hline
\end{tabular}
\end{center}
\end{table}
\newpage
\begin{table}
\begin{center}
\caption{Internal energy (K) of spin polarized liquid $^3{\rm He}$
versus total number density at $\xi=1/3$ computed with both
microscopic calculation and using parabolic approximation.}
\label{tab5}
\begin{tabular}{c|c|c}
\hline
 Density (${\rm A^{-3}}$) &Microscopic Calculation&Parabolic Approximation\\
\hline
 0.003 &       0.096&          0.127\\
 0.005 &      -0.268&         -0.212\\
 0.007 &      -0.688&         -0.611\\
 0.009 &      -1.135&         -1.100\\
 0.011 &      -1.532&         -1.477\\
 0.013 &      -1.444&         -1.420\\
 0.015 &      -0.523&         -0.517\\
 0.017 &       1.685&          1.695\\
 0.019 &       5.667&          5.697\\
\hline
\end{tabular}
\end{center}
\end{table}
\newpage
\begin{table}
\begin{center}
\caption{As Table~\ref{tab5}, but for spin polarized liquid
$^3{\rm He}$ at $\xi=2/3$.} \label{tab6}
\begin{tabular}{c|c|c}
\hline
 Density (${\rm A^{-3}}$) &Microscopic Calculation&Parabolic Approximation\\
\hline
 0.003&0.069&0.124\\
 0.005&-0.279&-0.164\\
 0.007&-0.652&-0.488\\
 0.009&-1.018&-0.867\\
 0.011&-1.382&-1.164\\
 0.013&-1.283&-1.186\\
 0.015&-0.349&-0.331\\
 0.017&1.874&1.895\\
 0.019&5.868&5.891\\
\hline
\end{tabular}
\end{center}
\end{table}
\newpage
\newpage
\newpage
\newpage
\newpage
\newpage
\end{document}